\documentclass[12pt]{article}   

\usepackage{amsfonts}   
\usepackage{amsmath}   
\usepackage{multicol}   
\usepackage{epsfig}   
\usepackage{graphicx}   

\def\a{\alpha} \def\b{\beta} \def\g{\gamma} \def\l{\lambda} \def\d{\delta} \def\e{\epsilon} \def\t{\theta} \def\r{\rho} 
\def\s{\sigma}    \def\m{\mu}   \def\p{\partial}  

\newcommand{\be}{\begin{equation}} 
\newcommand{\ee}{\end{equation}}
\newcommand{\bea}{\begin{eqnarray}} 
\newcommand{\eea}{\end{eqnarray}}

\begin{document}
 
  
\vbox{\hbox{YITP-SB-04-05}} 
\hbox{DAMTP-2004-04}
  
\vskip 1cm 
{\vbox{
\centerline{\bf \Large Superstrings and WZNW Models} 
}
 \vskip 1cm 
 
\medskip\centerline
{
P.~A.~Grassi$^{~a,b,}$\footnote{pgrassi@insti.physics.sunysb.edu},
G.~Policastro$^{~c,}$\footnote{G.Policastro@damtp.cam.ac.uk},  and  
P.~van~Nieuwenhuizen$^{~a,}$\footnote{vannieu@insti.physics.sunysb.edu}
} 
\vskip .5cm   
\centerline{$^{(a)}$ 
{\it C.N. Yang Institute for Theoretical Physics,} }  
\centerline{\it State University of New York at Stony Brook,   
NY 11794-3840, USA}  
\vskip .3cm  
\centerline{$^{(b)}$ {\it Dipartimento di Scienze e Tecnologie Avanzate,}} 
\centerline{\it Universit\`a del Piemonte Orientale} 
\centerline{\it C.so Borsalino, 54,  Alessandria, 15100 Italy}  
\vskip .3cm  
\centerline{$^{(c)}$ {\it DAMTP, Center for Mathematical Science, Wilberforce Road, } }  
\centerline{\it Cambridge CB3 0WA, UK}  

\medskip  
\vskip  .5cm  
\noindent  
We give a brief 
review of our approach to the quantization of superstrings. New is 
a covariant derivation of the measure at tree level and a path integral formula for this measure. 

  
\section{A brief review}

In a series of papers \cite{Grassione}
we have presented a new approach to the old problem of the covariant 
quantization of the superstring. The fields of the classical superstring are 
$x^{m}(z,\bar z)$ and two (i=1,2) spacetime spinors 
$\t^{\a i}(z,\bar z)$ where $m=0,\dots,9$ is a vector index and $\a =1, \dots, 
16$ the index of a real chiral spinor. A natural choice as action might seem to be  the supersymmetric line element 
\be\label{kin}
{S}_{1} = - {1\over 2}\int d^{2}x 
 \Pi^{m}_{\mu} \Pi^{\mu}_{m}\,, \ee 
where $\Pi^{m}_{\mu} = \p_{\mu} x^{m} - i \sum_{i} 
\t^{\a i} \g^{m}_{\a\b} \p_{\mu} \t^{\b i}$ 
with $\g^{m}_{\a\b} =\g^{m}_{\b\a}$ real Dirac matrices. 

It is invariant under the rigid supersymmetry transformations $\d_{\e} x^{m}
= i \e^{i} \g^{m} \t^{i}$ 
and $\d_{\e} \t^{\a i} = \e^{\a i}$. However, there is no kinetic term 
for $\t^{\a i}$. One can add a Wess-Zumino term \cite{GS}
\be
S_{WZ} = \int  \, \sum_{i} (-)^{i} d\t^{i} \g^{m} d\t^{i} \Pi_{m} \,, 
\ee 
It is separately supersymmetrically invariant, and $S_{1}+S_{WZ}$ has the (in)famous local 
$\kappa$ symmetry which allows one to choose the light-cone gauge in target space. 
Manifest Lorentz covariance  is then lost. So far the covariant quantization of this action has remained an open problem. 

Our approach begins with an idea of Berkovits \cite{berko}
 who started from the BRST charge 
 \be 
 Q_{B} = \oint dz \l^{\a} d_{\a} \,, ~~~~~~
 d_{\a} = p_{\a} + i \p x^{m} (\g_{m}\t)_{\a} + {1\over 2} (\g^{m} \t)_{\a} 
(\t \g_{m} \p \t)\,.
 \ee
The $\l^{\a}$ are commuting ghosts. 
We are now on the Euclidean worldsheet with light-cone derivatives 
$\p =\p_\s - i \p_{\tau}$ and restrict our attention to only one $\t^{\a}$. 
The OPE's $x^{m}(z) x^{n}(w) \sim - \eta^{mn} ln(z-w)$ and 
$p_{\a}(z) \t^{b}(w) \sim \delta_{\a}^{~\b} (z-w)^{-1}$ can be used 
to show that 
\be\label{i1}
d_{\a}(z)  d_{\b}(w) \sim 2 i {\g^{m}_{\a\b} \Pi_{m} \over (z-w)}.
\ee
The equation $d_{\a}=0$ yields 
the conjugate momentum of $\t^{\a}$ as obtained from the action $S_{1}+ S_{WZ}$ \cite{Grassione}, and is the only ingredient we retain of the Green-Schwarz model \cite{GS}. We build upon this $d_{\a}$ a whole new structure, and we expect that the results we obtain in the end will be equivalent to whatever one would call the covariant quantum superstring. 

From (\ref{i1}) it follows that $Q^{2}_{B} = 2 i \oint \l\g^{m} \l \Pi_{m}$. Clearly, 
if $\l$ are so-called pure spinors, which satisfy by definition $\l\g^{m}\l=0$, 
we obtain a nilpotent BRST charge \cite{berko}. These $\l^{\a}$ 
must be complex in order to have a nontrivial solution of $\l\g^{m}\l=0$. 

In \cite{Grassione} we 
have relaxed the constraint of pure spinors by adding a Lagrange multiplier term to $Q_{B}$ 
of the form $ - i \oint \b_{m} \l \g^{m}\l$. Now $\l^{\a}$ need not longer be complex, and we take it 
real, just like $\t^{\a}$.
In fact we first determined the superalgebra generated by $d_{\a}$; 
it contains in addition to $\Pi_{m}$ also $\p\t^{\a}$, and in the OPE's one finds 
terms with single poles and double poles. Denoting these three operators 
by $J^{g}_{M} = \{\Pi_{m}, - i d_{\a}, \p \t^{\a} \}$, the OPE has the form 
\be\label{i5}
J^{g}_{M}(z) J^{g}_{N}(w) \sim {J^{g}_{P} f^{P}_{~MN} 
\over (z-w)} - {{\cal H}_{MN}\over (z-w)^{2}}\,.
\ee
The BRST charge is then given 
\be
Q = - \oint dz ( J^{g}_{M} + {1\over 2} J^{gh}_{M} ) c^{M}
\ee
 where $c^{M} = \{\xi^{m}, \l^{\a}, \chi_{\a}\}$ are the 
ghosts, and the BRST ghost current $J^{gh}_{M}$ depends also on $b_{M}$, where 
$b_{M} = \{ \b_{m}, w_{\a}, \kappa^{\a} \}$ are the antighosts. The susy generator $q_{\a} = p_{\a} - i \p x^{m}(\g_{m}\t)_{\a} - {1\over 6} (\g_{m}\t)_{\a} (\t \g^{m} \p\t)$ transforms 
$\t^{\a}$ into $\e^{\a}$ and $x^{m}$ into $i \, \e\g^{m}\t$, and anticommutes 
with the covariant derivative $d_{\a}$.  

The three operators $J^{g}_{M}$ define a non-semisimple 
super Lie algebra \cite{gr}, but the invariant metric ${\cal H}_{MN}$ is still nonsingular. The 
generators of this algebra are $(P_{m}, Q_{\a}, K^{\a})$ where 
$K^{\a}$ is a fermionic central charge. In a coset approach $K^{\a}$ corresponds to new anticommuting coordinates $\phi_{\a}$ which form with the usual $\t^{\a}$ a bigger, but little 
understood superspace \cite{gr}. 

The ghost currents $J^{gh}_{M} = \{ 2 \kappa \g_{m} \l, 2 \xi^{m} (\g_{m}\kappa) + 
2 i \beta^{m} (\g_{m} \l, 0 \}$ satisfy the same OPE's as $J^{g}_{M}$ 
but without double poles. It then follows as usual that the  
BRST charge $Q$ is classically (using single contractions) nilpotent, 
but at the quantum level it ceases to be nilpotent due to the double poles. For that reason we have added, in the spirit of gauged WZNW models \cite{karabali}, 
a new triplet of currents $J^{h}_{M}$ which have the same OPE's as $J^{g}_{M}$ except 
that the sign of the double poles is opposite. The BRST charge is obtained by replacing $J^{g}_{M}$ by $J^{g}_{M}+ J^{h}_{M}$. Fatal double poles in contractions now cancel and this extended $Q$ is nilpotent. However, we have doubled all coordinates: in addition 
to $x^{m}, \t^{\a}$ and $p_{\a}$ we also have $x^{m}_{h}, \t^{\a}_{h}$ and $p^{h}_{\a}$. 

We have shown in \cite{Grassione} that a WZNW action based on the currents 
$J^{g}_{M}, J^{gh}_{M}$ and $J^{h}_{M}$, but in terms of $\phi_{\a}$ instead of $p_{\a}$, reduces to the free field action which in turn determines the propagators, provided one expresses $\p_{z} \phi_{\a}$ in terms of $p_{z \a}$ by $i \p_{z}\phi_{\a} = q_{z \a}$. Thus all our previous work turns out to have been based on a WZNW model. We work from now on in the formulation without $\phi_{\a}$ but with $p_{\a}$.  
 
This model has a very interesting conformal field theory. The BRST charge 
is 
\bea\label{BB}
j_{z}^{B} &=& - \l^{\a} (-i d_{z\a} + J^{h}_{z\a}) - \xi^{m}(\Pi_{z m}+ J^{h}_{z m}) 
- \chi_{\a}( \p_{z}\t^{\a} + \nonumber \\
&& + J^{h,\a}_{z}) - 2 \xi^{m}\kappa_{z}\g^{m} \l - i \b_{z m}(\l\g^{m}\l)\,.
 \eea
The energy-momentum tensor $T_{zz}$ is 
\bea
T_{zz} &=& - {1\over 2} \p_{z} x^{m} \p_{z} x_{m} - p_{z \a} \p_{z} \t^{\a} - 
\b_{z m} \p_{z} \xi^{m} - \kappa^{\a}_{z} \p_{z} \chi_{\a} - w_{\a} \p_{z} \l^{\a}
\nonumber \\
&&+ {1\over 2} J^{h}_{z m} J^{h, m}_{z} + i \, J^{h}_{z \beta} J^{h,\beta}_{z}  \,.
\eea
and the total conformal charge vanishes: 
$c=10-32 -20 +32 +32 +10 -32 =0$. The ghost current is
\be\label{gh}
j^{gh}_{z} = - \b_{z m} \xi^{m} - \kappa_{z}^{\a} \chi_{\a} -  w_{z \a} \l^{\a} \,.
\ee
Since the anomaly in the OPE $j^{gh}_z(z) j^{gh}_w(w) = c_j/(z-w)^2$ of the ghost current with itself is not zero but given by $c_j = - 22$, while $T_{zz}(z) j^{gh}_{w}(w) = 22/ (z-w)^{3} + j^{gh}_{z}(z)/(z-w)^{2}$, this superconformal algebra seems to be a twisted version of an 
$N=2$ superconformal algebra. The 
BRST current is nilpotent $j^{B}_{z}(z) j^{B}_{w}(w) \sim 0$. There is also 
a fermionic current $B_{zz}$ which is the dual of the BRST current
\be
B_{zz} = {1\over 2} (J^{g}_{M} - J^{gh}_{M}) b_{N} {\cal H}^{NM}\,.
\ee 
It has spin 2 and squares to $B_{zz}(z) B_{ww}(w) \sim {F_{www}(w)\over (z-w)}$.
The current $F_{zzz}$ is not only BRST closed but also BRST exact 
\bea
&&j^{B}_{z}(z) \Phi_{www}(w) \sim {F_{www} \over (z-w)}\,, \nonumber \\
&& \Phi_{zzz} = {-i \over 2} \b^{m}_{z} \kappa_{z} \g^{m} \kappa_{z} \equiv 
b_{Q} b_{P} b_{R} f^{RPQ} \,.
\eea
The six currents $j^{B}_{z}, B_{zz}, j^{gh}_{z}, T_{zz}, F_{zzz}$ and $\Phi_{zzz}$ generate 
a closed algebra. However, it is not an ordinarey N=2 superalgebra, but rather a deformation 
called a Kazama algebra \cite{kazama}. 

Following \cite{figue} one can recover 
an ordinary N=2 superconformal algebra by adding a topological gravity quartet, also 
called a Koszul quartet \cite{ver}. 
It contains the usual anticommuting ghost pair 
$(b_{zz}, c^{z})$ and a commuting spin $(2,-1)$ counterpart $(\b_{zz}, \g^{z})$. The ghost 
numbers are $(-1,1)$ and $(-2,2)$ respectively. The currents of this model form 
an ordinary $N=2$ superconformal algebra, so without $\Phi_{zzz}$ and $F_{zzz}$ 
currents. However, if one adds the currents of the Koszul model to the currents of the 
WZNW model, and modifies $B_{zz}$ appropiately, one ends up with an ordinary $N=2$ 
superconformal algebra for the combined system. 
 

\section{Tree level measure}

In this section, we give a derivation of the measure in our 
formalism. For that purpose, we follow the idea 
that the tree level action for the target space theory (the string field theory action) 
is determined by the so-called BV measure. Assuming that the target space action is of  
Chern-Simons type \cite{sft}, one has
\begin{equation} \label{2}
S_{D=10} = \int d\mu \left( \Phi Q \Phi + {2\over 3} \Phi \star \Phi \star \Phi 
\right) \,,
\end{equation} 
where $\Phi$ is the string field (see (13)) 
and the product $\star$ denotes the 
usual matrix multiplication and tracing over  $U(n)$ matrices. 
The integral in the above equation is the path 
integral over the zero modes of 
$x^{m}, \theta^{\a}, \lambda^{\a}, \xi^{m}$ and $\chi_{\a}$ (and $\xi$, see below)
which remains after one has performed the path over all nonzero modes. In order to 
perform the integration, we have to establish the measure $d\mu$. 
Since $\Phi$ has ghost number one, 
the ghost number of the Lagrangian is three. Therefore, 
we have to impose that the measure has ghost number $-3$
(notice that, in constrast to Berkovits' work we have 
more ghosts present, therefore there are several possible 
combinations). 

A second requirement is the BRST invariance of the measure.
To construct the measure it is convenient to consider a ghost number 
3 polynomial $\Omega^{(3)}$ which is invariant under super-Poincar\'e 
transformations (up to $Q$-exact terms) such that 
\begin{equation}\label{1}
\langle \Omega^{(3)} \rangle \equiv \int d\mu \, \Omega^{(3)} = 1\,.
\end{equation} 
The operator $\Omega^{(3)}$ represents the Poincar\'e dual to the 
measure (see for example Bott and Tu \cite{bott}). The condition of BRST 
invariance of  $\Omega^{(3)}$ is therefore given by $\int d\mu 
\{Q, \Omega^{(3)}\} = 0$. 

The measure $d\m$ itself should also be BRST invariant. We implement this 
requirement by imposing 
the condition that if  $\Omega^{(3)} = \{Q, {\cal K} \}$ where 
${\cal K}$ is any polynomial with ghost number 2, then 
\be\label{cic2}
\int d\mu  \{Q, {\cal K} \} = 0 \,.
\ee 

Since the action in (\ref{1}) should reproduce the 
super-Yang-Mills action in $d=(9,1)$ dimensions, we 
can choose a given coupling to normalize the integral in (\ref{1}). 
For example the gluon-gluino-gluino coupling 
$f_{abc} v^{a}_{\mu}(x) \psi^{b\a} \g^{m}_{\a\b} \psi^{c \b}$ may provide a convenient normalization. A generic vertex reads 
\bea\label{3}
\Phi &=& \lambda^{\a} A_{\a} + \xi^{m} A_{m} + \chi_{\a} W^{\a} + \nonumber \\
&& \vspace{-2cm}
+ \xi \left( \xi^{m} \xi^{n} F_{mn} + \xi^{m} \chi_{\a} F_{m}^{~\a} + 
\lambda^{\a} \chi_{\b}  F_{\a}^{~\b} + \chi_{\a} \chi_{\b} F^{\a\b} \right)\,,
\eea
where the superfields $A_{\a}, \dots, F^{\a\b}$ depend 
only the combinations $x^{m}-x^{m}_{h}$ and so on. (The other 
combinations are excluded by imposing the condition $B_{0}\Phi =0$ 
on the vertex operator.) We also assume that all superfields are Lie algebra 
valued.  The field $\xi$ is obtained by the bosonization of the superghosts 
$\gamma^{z} = \eta e^{\phi}$ and $\beta_{zz}  = \p\xi e^{-\phi}$ of the 
twisted topological Koszul quartet. 

By expanding the superfields $A_{m}$ and $W^{\a}$ one 
finds that the first coefficients coincide with the gluon field $v^{a}_{m}$ and 
with the gluino $\psi^{a \a}$, respectively. Therefore, a candidate for $\Omega^{(3)}$ is 
given by 
\begin{equation}\label{4}
\Omega^{(3)} \sim \g^{m, \a\b} \chi_{\a} \xi_{m} \chi_{\b}\,.  
\end{equation}
because substitution into (11) and (10) reproduces the gluon-gluino-gluino coupling. 
Notice that this candidate for $\Omega^{(3)}$ has the correct 
ghost number, and also from a conformal point of view it has 
the right properties, namely it is a scalar, just like the corresponding 
element $\Omega^{(3)}_{bos} = c \p c \p^{2} c$ 
of the bosonic string.  
The form of $\Omega^{(3)}$ and equation (\ref{1}) imply that 
\begin{eqnarray}\label{4.1}
&&\langle (\xi^{m} A_{m})(z_{1}) 
(\chi_{\a} W^{\a})(z_{2}) (\chi_{\a} W^{\a})(z_{3}) \rangle \rightarrow  \\
&&~~~~~~~
\rightarrow \int d^{10}x \, 
f_{abc} v^{a}_{m} \psi^{b \a} \psi^{c \b} \langle \xi_{m} \chi_{\a} \chi_{\b} \rangle = 
\int d^{10}x f_{abc}  v^{a}_{m} \psi^{b \a} \psi^{c \b}\, \g^{m}_{\a\b}\,. \nonumber
\end{eqnarray}

However, $\Omega^{(3)}$ in (\ref{4}) fails to be BRST invariant. 
To repair it, we consider the following two combinations of ghosts 
\begin{equation}\label{5}
\hat \xi^{m} = \xi^{m} +  \l^{\a} \g^{m}_{\a\b} \t^{b}\,, ~~~~~
\hat \chi_{\a} = \chi_{\a} - 2 i \xi^{m} \g_{m\a\b} \t^{\b} + {4 i \over 3}
\g^{m}_{\a\b} \t^{\b} \l^{\g} \g_{m\g\d} \t^{\d}
\end{equation} 
which are invariant under BRST transformations. Then 
we have finally
\begin{eqnarray}\label{6}
\Omega^{(3)} &=&   \g^{m}_{\a\b} \hat\chi_{\a} \hat\xi_{m} \hat\chi_{\b} \\
&=& (\chi \g^{m} \chi) \xi_{m} + 2 \xi^{m} \xi^{n} \chi \g_{mn}\t + 
\xi^{m} (\chi\g_{m} \g_{n} \t) ( \l \g^{n} \t)  + (\l \g^{n} \t) (\chi \g_{n} \chi) \nonumber \\
&+& (\l\g^{m}\t)  (\l\g^{n}\t)  (\chi \g_{mn} \t) + 
\xi^{m}\xi^{n} \xi^{r} \t\g_{mnr}\t + 
\xi^{m} \xi^{n} (\l\g^{r}\t) \t\g_{mnr}\t  \nonumber \\
&+&
\xi^{m} (\l\g^{n}\t) (\l\g^{r}\t)\t\g_{mnr}\t + 
(\l\g^{m}\t)(\l\g^{n}\t)(\l\g^{r}\t) \t\g_{mnr}\t  
\nonumber 
\end{eqnarray}
where the first term reproduces the combination needed to fix the coupling 
of the gluon-gluino-gluino vertex. More important is the 
observation that the last term in  $\Omega^{(3)}$ is exactly 
the combination discussed in \cite{berko}. Substitution of this term with pure spinors into 
(12) and (11) reproduces the complete super-Yang-Mills theory, and the measure is in this case
also BRST invariant. In our covariant approach the $\l$'s are not constrained to be pure spinors, and then we need all terms in (\ref{6}) for BRST invariance. 

In order, to 
fix the integration, one has to assume that a particular 
monomial $\Omega^{(3)}_{i}$ 
of $\Omega^{(3)}$ has the property $\int d\mu \Omega^{(3)}_{i} = 1$. 
One choice is obviously the last term, as we have learnt from \cite{berko}. 
Another choice is to fix the first term. However, we can also 
impose that  another combination of terms gives the expected result.  

\section{Path Integral formula for the measure}

In order to show that the ghosts and antighosts of the present formulation 
are emerging from the Berkovits formulation, we derive here a path integral formula 
for the measure (\ref{6}) assuming that only the last piece in (\ref{6}) (the 
so-called Berkovits term) 
$\Omega^{(3)} = (\l\g^{m}\t)(\l\g^{n}\t)(\l\g^{r}\t) \t\g_{mnr}\t$ contributes to (\ref{1}). 
We shall discuss in section 4 that this is indeed the case, but that at the same 
time we can continue with 16 unconstrained and real $\l^{\a}$. 
Therefore, we write
\be\label{p1}
1 = \int d^{16}\t \, d^{16}\l \, \m(\l,\t) \, [(\l\g^{m}\t)(\l\g^{n}\t)(\l\g^{r}\t) \t\g_{mnr}\t] \,.
\ee
and we want to compute $\mu(\l,\t)$. 
At the first step, we notice that  in order that the integral over the pure spinor fields 
$\l$ is well defined the measure should contain a Dirac delta function to render the integration over the $\l$'s finite: $\mu(\l,\t) = \delta^{16}(\l) \mu(\l, \t)'$. The delta function in the 
measure is also needed for compensating the ghost number of $d^{16}\l$. In order that 
the measure $\mu'(\l,\t)$ projects out three components of the pure spinor, 
it should have the form
\be\label{p2}
\mu(\l, \t) =  \delta^{16}(\l) \mu^{(\a\b\g)} (\theta) \p_{\l^{\a}} \p_{\l^{\b}} \p_{\l^{\g}}\,. 
\ee
One needs exactly three $\l$-derivatives to take into account the ghost number. 

The next step is to represent the 
delta function by an integral over variables 
$w_{\a}$ which we identity with our antighosts.\footnote{This needs further study because the antighosts $w_{z\a}$ are vectors on the worldsheet which do not have zero modes on a genus zero Riemann surface.} The three derivatives become then three components of $w$
\be\label{p3}
1 = \int d^{16}\t \, d^{16}\l \, d^{16}w \, e^{i w_{\a} \l^{\a}} 
\, [\m^{(\a\b\g)}(\t) w_{\a} w_{\b}w_{\g}] \, \Omega^{(3)} \,.
\ee
The 
integration $d^{16}\t$ should be saturated by the $\t$'s 
present in $\m^{(\a\b\g)}(\t) $ and those present in $\Omega^{(3)}$. 
Knowing that in  the latter there are 5 $\t$'s, we obtain  
\be\label{p4}
\m^{\a\b\g}(\t) = \m^{(\a\b\g)}_{[\r_{1} \dots \r_{11}]} \t^{\r_{1}} \dots \t^{\r_{11}}\,,
\ee
where $\m^{(\a\b\g)}_{[\r_{1} \dots \r_{11}]}$ are numerical constants. Using the 
following formula for Berezin integrals 
\be\label{p5}
\t^{\a_{1}} \dots \t^{\a_{n-l}} = 
{ (n-l)! (-i)^{l} \over l!} \e^{\a_{1} \dots \a_{n-l} \r_{n-l+1} \dots \r_{n}} 
\int d^{n}p  \left(p_{\r_{n-l+1}} \dots p_{\r_{n}} \right) e^{i p_{\a} \t^{\a}}\,,
\ee
 where $p_{\a}$ are 16 fermionic new variables and $\e^{\a_{1} \dots \a_{16}}$ is 
 the invariant tensor of $Spin(9,1)$, we obtain
 \be \label{p6}
 1 = \int d^{16}\t \, d^{16}\l \, d^{16}w e^{i \l^{\a} w_{\a} + i p_{\a} \t^{\a}}
  \, [(p \g^{m} w) \, (p\g^{n} w) \, (p \g^{r}w_{\g})\,  (p\g_{mnr} p)] \, \Omega^{(3)} \,.
\ee
The contractions in (\ref{p6}) 
with the invariant tensors are dictated by Lorentz invariance. In particular we have 
\be\label{p7}
\m^{(\a\b\g)}_{[\r_{1} \dots \r_{11}]} =  
\e_{\a_{1} \dots \a_{5} \, \r_{1} \dots \r_{11}} 
\g^{m, \a \a_{1}} \g^{n, \b \a_{2}} \g^{r ,\g \a_{3}} \g_{mnr}^{\a_{4} \a_{5}}\,. 
\ee

To simplify further the integral, we introduce 10 anticommuting 
variables $\xi^{m}$ which correspond exactly to our ghost fields 
$\xi^{m}(z)$ discussed in the previous sections. We can then rewrite 
(\ref{p6}) as follows
\bea \label{p8}
1 &=& 
\int d^{16}\t \, d^{16}\l \, d^{16}w d^{10} \xi e^{i \l^{\a} w_{\a} + i p_{\a} \t^{\a} + 
\xi^{m} p \g_{m} w} \times \nonumber \\
&& \times 
 \, \left[ \e_{m_{0} \dots m_{9}} \xi^{m_{0}} \dots \xi^{m_{6}}
  (p\g^{m_{7} m_{8} m_{9}} p)\right] \, \Omega^{(3)} \,.
\eea
Again it is convenient to use (\ref{p5}) to introduce variables $\b_{m}$ 
which correspond to our antighosts $\b^{m}_{z}(z)$ and which simplify 
the expression for the measure further. This leads to 
\bea\label{p9}
1 
&=& 
\int d^{16}\t \, d^{16}p\,  d^{16}\l \, d^{16}w \, d^{10} \xi \, d^{10} \b
\times \nonumber \\
&& \times 
 e^{(i \l^{\a} w_{\a} + i p_{\a} \t^{\a} + i \xi^{m} \b_{m}+  
\xi^{m} p \g_{m} w)} 
  \, \left(p\!\not\!\b\!\not\!\b\!\not\!\b \, p\right) \, \Omega^{(3)} \,. \nonumber 
\eea
where $\not\!\b_{\a\b} = \g^{m}_{\a\b} \b_{m}$. As a last step
we introduce commuting variables $\chi_{\a}$ and $\kappa^{\a}$ which 
are the final pairs of (anti)ghosts in our work. 
The operator $\left(p\!\not\!\b\!\not\!\b\!\not\!\b \, p\right)$ can be written 
as follows
\be\label{p10}
\left(p\!\not\!\b\!\not\!\b\!\not\!\b \, p\right) = 
\p_{\chi_{\a}} \left(p\!\not\!\b \chi\right) \g^{m}_{\a\b} \b_{m} \, \p_{\chi_{\a}} 
\left(p\!\not\!\b \chi\right) \,,
\ee
The integration over $\chi_{\a}$ requires the introduction of another delta 
function $\d^{16}(\chi)$ in order to make these integrals finite. 
So we have
\bea\label{p11}
1 &=& 
\int d^{16}\t \, d^{16}p \, 
d^{16}\l \, d^{16}w \, 
d^{10} \xi \, d^{10} \b \, 
d^{16}\chi \, d^{16}\kappa 
\times \nonumber \\
&& \times 
e^{(i \l^{\a} w_{\a} + i p_{\a} \t^{\a} + i \xi^{m} \b_{m}+i \kappa^{\a} \chi_{\a}+   
\xi^{m} p \g_{m} w + \b_{m} p \g^{m} \chi)} 
\left( \kappa^{\a} \g_{\a\b}^{m} \kappa^{\b} \b_{m} \right) 
\, \Omega^{(3)} \,. \nonumber 
\eea

We have thus constructed a formula for the measure 
\bea\label{p12.0}
\m(\t, \l; \xi, \b, \chi, \kappa, w, p) &=& \nonumber \\ 
&& \hspace{-3cm} 
e^{(i \l^{\a} w_{\a} + i p_{\a} \t^{\a} + i \xi^{m} \b_{m}+i \kappa^{\a} \chi_{\a}+   
\xi^{m} p \g_{m} w + \b_{m} p \g^{m} \chi)} 
\left( \kappa^{\a} \g_{\a\b}^{m} \kappa^{\b} \b_{m} \right) \,,
\eea
which contains the Fourier transform of all independent pairs 
$(\l,w), \dots (\kappa, \chi)$. The remaining piece 
$\left( \kappa^{\a} \g_{\a\b}^{m} \kappa^{\b} \b_{m} \right)$ has a very interesting 
property. It can be identified with the spin 3, ghost number -3 field 
$\Phi_{zzz}(z) = \left( \kappa^{\a}_{z} \g_{\a\b}^{m} \kappa^{b}_{z} \b_{z m} \right)$ appearing in the previous section. Notice that we can also make a redefinition of the 
fields $\l^{\a} \rightarrow \l^{\a} - i  \xi^{m} (\g_{m} p)^{\a}$ and $\kappa^{\a} \rightarrow 
\kappa^{\a} - i \b^{m} (\g_{m} p)^{\a}$ to recast the measure in a more elegant form
\bea\label{p12}
\m(\t,\l; \xi, \b, \chi, \kappa, w, p) = 
e^{(i \hat\l^{\a} w_{\a} + i p_{\a} \t^{\a} + i \xi^{m} \b_{m}+i \hat\kappa^{\a} \chi_{\a})} 
\left( \kappa^{\a} \g_{\a\b}^{m} \kappa^{\b} \b_{m} \right) \,.
\eea 

We conclude that we have provided a path integral formula for the tree level measure. 
We need all ghosts of 
our formalism to be able to construct the formula (\ref{p12}). 
We have recovered the 
operator $\Phi_{zzz}$ from a completely different point of view. 
The exponent in (\ref{p12}) suggests that the ghost current should be 
changed to $j^{gh} \rightarrow j^{gh} + p_{z\a} \t^{\a}$. Notice that the action 
$\int d^{2}z p_{z\a} \bar\p \t^{\a}$ is invariant under the rigid transformations 
generated by $\oint p_{z\a} \t^{\a}$. The new ghost current $j^{gh}$ has an anomaly 
equal to $c_{j} = 6$. 
 
\section{Summary and Open Problems}

We have constructed a tree level measure for the string field 
theory formulation of our approach to the superstring. As we shall publish elsewhere, one can introduce two further BRST charges, such that all terms 
in (18) except the last one, are BRST exact. In the vertex operator $\Phi$ 
one may then drop all terms except $\l^{\a}A_{\a}$. 
Then we only need retain the following terms 
\be 
A_{\a} \sim a_{m} (\g^{m}\t)_{\a} + (\g_{m}\t)_{\a} (\t\g^{m}\psi) + F_{mn} (\t\g^{mnt}\t) 
(\g_{t}\t)_{\a} + {\cal O}(\t^{5})\,. 
\ee
(These are the only structures which remain in the WZ gauge $\t^{\a}A_{\a} =0$.) 
The action in (10) then yields the complete SYM action in 10 dimensions. 

So the following picture emerges. Our completly covariant approach with many more 
ghosts then only $\l^{\a}$ has linearized the theory to a point where no constraints 
are any longer present, and no non-covariant canonical approach is needed. Yet, 
at the end we get rid of all excess ghosts, and recover the formulation with only 
$\l^{\a}$. Thus this approach combines the best of the covariant approach with 
the maximal number of ghosts and the minimal approach with only $\l^{\a}$. We hope 
that these ideas will also be effective at the loop level. 
\vspace{.2cm}

There are (of course) many problems left: 
\begin{itemize}
\item To define physical states we had to 
restrict vertex operators such that they have ``non-negative grading''. The grading is a 
quantum number one can assign to the ghosts because the underlying algebra is non-semisimple. In \cite{Grassione} 
we have shown that it is related to homological perturbation theory, but it is 
desirable to find an extra BRST operator which selects non-negative grading. 

\item 
We should calculate the cohomology of the model with $J^{g}_{M}, J^{gh}_{M}$ and 
$J^{h}_{M}$ together with the Koszul quartet. We have found two BRST 
charges which anticommute, and further conditions involving $B_{zz}(z)$ which remove the dependence on the differences $x^{m}-x^{m}_{h}$, etc, but a complete analysis is in preparation. 

\item So far we focused only on the left-moving sector {\it i.e.} the 
heterotic string, but a complete treatment containing 
both left-moving and right-moving 
sectors for all fields, in particular for the WZNW model, is still lacking. Also 
this we intend to construct. 

\item The big problem is, of course, the calculation of amplitudes: path integrals, measures and 
vertex operators. We hope that here our covariant methods will be helpful. 
\end{itemize}


\section*{Note added}

On the day we submitted this article to the hep-th arXive we heard a seminar by N. Berkovits, who constructed  a general expression for the measure in his approach with pure spinor constraints. Although there are differences between his and our approach, for example complex vs. real $\l^{\a}$ ghosts resulting in integrations over 11 instead of 16 $\l$'s, there is 
also overlap with the present work. 


 \end{document}